\documentclass[12pt]{article}
\def\slash#1{\setbox0=\hbox{$#1$}#1\hskip-\wd0\hbox to\wd0{\hss\sl/\/\hss}}
\usepackage{epsf, cite, amsmath, amssymb}
\usepackage{epsfig}
\usepackage[all]{xy}
\makeatletter
\renewcommand\section{\@startsection {section}{1}{\z@}%
                                   {-3.5ex \@plus -1ex \@minus -.2ex}
                                   {2.3ex \@plus.2ex}%
                                   {\normalfont\large\bfseries}}
\renewcommand\subsection{\@startsection{subsection}{2}{\z@}%
                                     {-3.25ex\@plus -1ex \@minus -.2ex}%
                                     {1.5ex \@plus .2ex}%
                                     {\normalfont\bfseries}}
\makeatother

\newcommand{\bea}{\begin{eqnarray}}
\newcommand{\eea}{\end{eqnarray}}
\newcommand{\be}{\begin{equation}}
\newcommand{\ee}{\end{equation}}

\newcommand{\e}{\epsilon}

\newcommand{\p}{\partial}

\newcommand{\s}{\sigma}
\newcommand{\del}{\delta}

\def\Pr{ { {\cal{P}}_{\cal{R}} }  }
\def\Prp{ { {\cal{P}}_{\cal{R^+}} }  } 
\def\Prm{  {\cal{P}}_{ {\cal{R^-}} }  }
\def\cRp{ { { \cal{R}}^+}  } 
\def\cRm{ { {\cal{R}}^-}  }
\def\cR{ {\cal{R}}  }

\newcommand{\C}[1]{$(\ref{#1})$}

\begin{document}

\begin{titlepage}

\begin{center}

\hfill EFI-04-41

\vskip 2 cm
{\Large \bf The M2-M5 Brane System and a Generalized Nahm's Equation}\\
\vskip 1.25 cm { Anirban Basu\footnote{email: basu@theory.uchicago.edu}
and Jeffrey A. Harvey\footnote{email: harvey@theory.uchicago.edu}
}\\
{\vskip 0.75cm
Enrico Fermi Institute and Department of Physics, University of Chicago,\\
5640 S. Ellis Avenue, Chicago IL 60637, USA}

\end{center}

\vskip 2 cm

\begin{abstract}
\baselineskip=18pt

We propose an equation that describes M2-branes ending on M5-branes,
and which generalizes the description of the D1-D3 system via Nahm's equation.
The simplest solution to this equation constructs the transverse geometry
in terms of a fuzzy three sphere. We show that the solution passes a number
of consistency checks including
a calculation of the energy of the system, matching to the self-dual 
string solution in the M5-brane world volume, and a study of simple 
fluctuations
about the ground state configuration.
We write down certain terms in the effective action of multiple 
membranes, which includes a sextic scalar coupling.

\end{abstract}

\end{titlepage}

\pagestyle{plain}
\baselineskip=19pt

\section{Introduction}
It is possible to have branes ending on branes in both string and M 
theory~\cite{Strominger:1995ac,Townsend:1995af,Townsend:1996em}. In M theory,
the boundary of an M2-brane ending on an M5-brane describes a BPS self--dual 
string soliton--the classical soliton solution to the five brane equations 
of motion has been constructed in~\cite{Howe:1997ue}. Our aim is to
try to analyze the membrane theory and understand the geometry
when multiply coincident M2 branes end on a five brane.

Similar constructions have been made in string theory. One can have 
D-strings ending on D3-branes, and these BPS saturated Bion solutions 
have been 
constructed in~\cite{Callan:1997kz,Gibbons:1997xz}, as spikes in the world 
volume theory on the D3-branes.\footnote{Similar BPS bounds for 
dyons have been constructed 
in~\cite{Gauntlett:1997ss,Brecher:1998tv}.} The boundary of $k$ 
coincident D-strings ending on $N$ D3-branes can be interpreted as a $k$ 
monopole solution in the $SU(N)$ gauge theory. From the point of view of the 
D-string world volume theory, the corresponding Bogomolnyi equation turns 
out to be the Nahm equation~\cite{Nahm:1979yw} for the moduli 
space of monopoles in the gauge theory~\cite{Diaconescu:1996rk} (see 
also~\cite{Kapustin:1998pb,Tsimpis:1998zh,Giveon:1998sr}).

One can construct solutions to the D-string equations of motion which open 
up into a funnel representing the D3-brane~\cite{Constable:1999ac}, 
obtaining a match between the D3 and the D1 points of view. Here the 
transverse coordinates of the D-strings representing the D3 world volume 
define a noncommutative (fuzzy) two sphere~\cite{Madore:1991bw}.
(D0-branes when placed in a four form field strength background expand into
a fuzzy two sphere as well~\cite{Myers:1999ps}) .
Similar solutions for D-strings opening into D5-branes has been 
constructed~\cite{Constable:2001ag} with the transverse coordinates 
forming a fuzzy four sphere~\cite{Grosse:1996mz,Castelino:1997rv}. 
(For D-strings opening into D7-branes see~\cite{Cook:2003rx}.) 

Our aim is to understand something similar for the case of multiple 
membranes ending on a fivebrane in M theory. The strategy is to construct 
``solutions'' to the membrane world volume theory which represent 
this configuration and obtain a funnel--like 
solution representing a five brane growing out of the membranes. However, 
the world volume theory of multiple membranes is not known. So we proceed by
using various analogies with D-brane systems, matching the solution
with the soliton solution in~\cite{Howe:1997ue} and making some
consistency checks.

It should be noted that the matrices we describe here are very different 
from the ones in Matrix theory~\cite{Banks:1996vh}. In Matrix theory, the 
large $N$ limit of the matrices gives a single membrane, while in our case we 
represent the degrees of freedom by $N$ by $N$ matrices.

It is not entirely clear to us whether these matrices should be thought of
as the fundamental variables of the ultraviolet description of 
multiple membranes,
or some effective description which happens to capture the behavior of the
M2-M5 brane system.

In the next section we remind the reader of Nahm's equation and the D1-brane
description of the D3-D1 system.
We then propose a generalization to the M5-M2 system and subject our proposal
to a number of consistency checks, including matching to the self-dual string
solution on the M5-brane, an analysis of 
the energy of the system, and
a study of  a particular mode representing transverse fluctuations of 
the membranes which we show to be consistent with the interpretation of 
this configuration as membranes ending on an M5-brane. We end with a summary
and discussion of open problems. 

\section{Review of the D1-D3 system}

Our proposal for the equation describing self-dual string solutions is 
based on a generalization
of Nahm's equation for magnetic monopoles. This in turn arises as the 
description of the D1-D3
brane system from the point of view of the D1 world-volume theory 
\cite{Diaconescu:1996rk}. In this section we review the
main facts about the D1-D3 system that we want to generalize in our 
description of the M2-M5 brane
system.

We first consider $N$ coincident D1-branes ending on a single D3-brane. 
The D3 world-volume
description of this configuration was found in \cite{Callan:1997kz},
\cite{Howe:1997ue}, \cite{Gibbons:1997xz}. We take
the D3 world-volume to lie along the directions 0,1,2,3 and the D1 
world-volume to lie along 0,9. The
solution corresponding to the stack of N D1-branes excites a magnetic 
field on the D3-brane
and one of the transverse scalars with a profile given by
\be
X^9 = N \pi \alpha'  /\sqrt{(X^1)^2 + (X^2)^2 + (X^3)^2}.
\ee

>From the D1 world-volume point of view the solution has three transverse 
scalars $X^i$, $i=1,2,3$
excited as a function of the spatial world-volume coordinate $X^9$ and 
obeys the Nahm equation
\be
{\partial X^i \over \partial X^9} \mp {i \over 2}\epsilon^{ijk} [X^j,X^k] = 0.
\ee
This can be solved by 
\be
X^i = \pm {1 \over 2 X^9} \alpha^i,
\ee
provided that the $\alpha^i$ obey the Lie algebra of $SU(2)$:
\be
[ \alpha^i,\alpha^j ] = 2 i \epsilon^{ijk} \alpha^k.
\ee

Choosing the $\alpha^i$ to be in the $N$-dimensional irreducible 
representation of
$SU(2)$ with quadratic Casimir $C= N^2 -1$, the solution, at fixed 
$X^9$, describes a
fuzzy or non-commutative two sphere with physical radius
\be
R = \sqrt{ {(2 \pi \alpha')^2 \over N} \sum_i Tr [ (X^i)^2]} = {(2 \pi 
\alpha') \sqrt{ (N^2-1)} \over 2 X^9}.
\ee
At large $N$ this matches exactly with the result derived from the D3 
world-volume point of
view.

In \cite{Constable:1999ac} other evidence is given that the D1 world-volume 
description gives
a correct description of the D1 ending on a D3-brane. This includes matching 
of the energy
of the configuration, and the existence of fluctuations in the overall 
transverse directions which
fluctuate along the D1 and then out into the D3 and obey the correct 
higher-dimensional
wave equation. 

We now turn to our proposed generalization of Nahm's equations and the 
description of the
M2-M5-brane juncture from the M2 world-volume point of view. A different
modification of Nahm's equation relevant to the D2-D4 system has been
studied in \cite{Campos:2000de}.

\section{The M2-M5 system}

\subsection{A generalized Nahm's equation}

We first recall the description of $N$ M2-branes ending on an M5-brane 
from the M5
world-volume point of view.  This is given by the self--dual string 
soliton constructed in~\cite{Howe:1997ue}. This solution has one 
collective coordinate 
on the five brane world volume theory excited--it is the direction along 
which the membranes extend away from the five brane. If $s$ parametrizes 
this direction, then
\be \label{sdep} s \sim \frac{N}{R^2},\ee  
where $R$ is the distance in the five brane world volume. So this 
collective coordinate represents a ``ridge'' solution in the 
M5-brane theory. 
This solution is valid for large $R$, when the fields in the M5-brane 
theory are slowly varying. We now construct an ansatz for the membrane 
world volume theory valid near the core of the ridge (large $s$) which 
satisfies this condition. We shall later explain the validity of matching 
the solutions which are naively valid in different regimes.   

To generalize Nahm's equation to this situation we want an equation
which has $SO(4)$ symmetry (rather than the $SO(3)$ symmetry of Nahm's
equation), which has translation symmetry, which constructs the 
M5-brane using a fuzzy construction of the
three-sphere, and which gives the correct result for the physical 
radius of the
M2-M5 brane configuration as determined by the self-dual string 
soliton solution.

The construction of a fuzzy three-sphere is discussed in Appendix A
following the original construction in ~\cite{Guralnik:2000pb}. 
The algebra is constructed 
in terms of representations of $spin(4)
=SU(2) \times SU(2)$, $\cRp, \cRm$  given by the 
$(\frac{n+1}{4},\frac{n-1}{4})$ and $(\frac{n-1}{4},\frac{n+1}{4})$ 
representations respectively with $n$ an odd integer. The dimension of
$\cR = \cRp \oplus \cRm$ is $N=(n+1)(n+3)/2$. The coordinates on the 
fuzzy $S^3$
are $N \times N$ matrices $G^i$ which map $\cRp$ to $\cRm$ and $\cRm$ to
$\cRp$. The matrix $G_5$ is given in terms of the
difference of projection operators onto these representations,
\be G_5 = \Prp - \Prm. \ee

As discussed in ~\cite{Ramgoolam:2001zx}, there are three closely related
algebras related to the $G^i$. One is the algebra generated by taking arbitrary
products of the $G^i$. The second is the algebra of $N$ by $N$ complex
matrices, $Mat_N(C)$, which contains the first algebra as a subalgebra. 
Finally,
there is a projection of the Matrix algebra which in the large N limit agrees
with the classical algebra of functions on $S^3$. It is not entirely clear
to us which of these three algebras should be used in our construction. For
the purposes of this paper we use $Mat_{N}(C)$ but keep in mind the possibilty
that future developments may require a refinement of this choice. 

Taking the matrix M2-coordinates to be in $Mat_{N}(C)$, our proposed
equation is then
\be \label{nahm}
\frac{d X^i}{ds} +  
\frac{\lambda M_{11}^3}{8\pi} \e_{ijkl}
{1 \over 4!}[G_5, X^j, X^k, X^l] =0,\ee
where the quantum Nambu 4-bracket is defined by
\be [A_1,A_2,A_3,A_4] = \sum_{{\rm permutations} 
~ \sigma} {\rm sgn}(\sigma) A_{\sigma_1}
A_{\sigma_2} A_{\sigma_3} A_{\sigma_4}, \ee
and $\lambda$ is an arbitrary parameter which we fix shortly.
This structure has appeared in other attempts to define an odd quantum
Nambu bracket
\cite{Nambu:1973qe},\cite{Curtright:2003wj}, \cite{Sheikh-Jabbari:2004ik}.

Actually, for the solutions we will discuss, all $4!$ terms in the 4-bracket
are equal, and so there are many equivalent ways that one could write the
equation. The 4-bracket has the most natural mathematical form, but future
developments may prefer some other form of the equation which is equivalent
on the solution we describe below.

The equation \C{nahm} is manifestly $SO(4)$ invariant as a result of 
the $SO(4)$
action on the fuzzy $S^3$. It is also clearly invariant under translations 
taking
$X^i \rightarrow X^i + v^i I$ with $I$ the identity matrix.

The equation \C{nahm} should be the Bogomolnyi equation 
for the membrane theory
and should follow from the vanishing of the supersymmetry variation of 
the fermion
fields, that is from the BPS condition. 

We will study the equation \C{nahm} on the semi-infinite interval
$s \in (0,\infty)$ which is appropriate to semi-infinite M2-branes
terminating on an M5-brane. In analogy to Nahm's equation, it would also
be interesting to study this equation on the finite interval $s \in (-1,1)$
which would correspond to finite M2-branes suspended between M5-branes.
In this case the discussion below suggests that the proper boundary
conditions are that
\be
X^i (s) \sim \frac{G^i}{\sqrt{s \pm 1}},
\ee
where the $G^i$ are defined in Appendix A.

\subsection{Comparison to the Self-Dual String Solution}

We now construct a solution of \C{nahm} that represents $N$ M2-branes
ending on a single M5-brane.
To do this we make an ansatz for $X^i$ in terms of a radial function 
and the 
coordinates on the fuzzy three sphere defined in Appendix A.
\be \label{defX} 
X^i (s)= \frac{i{\sqrt{2\pi}}}{\sqrt{\lambda (n+2)} M_{11}^{\frac{3}{2}}} f(s)
G^i,\ee
where $M_{11}$ is the eleven dimensional Planck mass.  Then, 
using the identities in Appendix
A, it is easy to see that this solves \C{nahm} in the large $N$ 
limit provided that
\be
f(s) = {1 \over \sqrt{s}}.
\ee
Sending $X^i \rightarrow G_5 X^i$, we get 
\C{nahm} with a relative minus sign between the two terms, describing the 
solution for anti--self dual strings.

We can also 
define a distance
$\hat{R}$ given by
\be \hat{R} = \frac{{\sqrt{2\pi}}}{\sqrt{\lambda (n+2) s} 
M_{11}^{\frac{3}{2}}},\ee
such that $X^i = i \hat{R} G^i$. We also define the physical radius $R$ by
\be \label{defR}
R = {\sqrt {\frac{1}{N} \vert {\rm Tr} \sum_i (X^i)^2 \vert } } = 
{\sqrt N} \vert \hat{R} \vert.\ee 
We would like the quantities $s$ and $R$ in \C{defX} and \C{defR} to have 
the same interpretation as the ones in \C{sdep}. Now \C{defR} 
yields 
\be s= \frac{2\pi N}{\lambda (n+2) M_{11}^3 R^2},\ee
which is not of the form \C{sdep} for arbitrary $\lambda$. 

We shall soon see that 
our analysis is valid for large $N$, and  in this limit we will have the same scaling
with $R$ and $N$ provided that $\lambda^2 N $ is held fixed at large $N$. 
Since $\lambda$ can be interpreted as the coupling in the theory (which will turn out to be 
classically marginal from the discussion below), this amounts to an analogue of
the 't Hooft coupling being held fixed at large $N$.

\subsection{Energy and Action}

Given the Nahm equation, it is natural to define 
the energy of this static configuration by the expression
\bea \label{energyexp}
E = T_2 \int d^2 \sigma {\rm Tr} \bigg[ \left( \frac{d X^i}{ds} + 
\frac{\lambda M_{11}^3}{8\pi} \e_{ijkl} 
G_5 X^j X^k X^l \right)^2  \cr 
+ \left(1- \frac{\lambda M_{11}^3}{16\pi} \e_{ijkl} \left\{ \frac{d X^i}{ds} ,
G_5 X^j X^k X^l \right\} \right)^2 \bigg]^{1/2} ,\eea
where $T_2  = M_{11}^3 /(2\pi)^2$ is the membrane tension and we have 
integrated over the spacelike membrane worldvolume directions, where
$\sigma_1=\sigma$ and $\sigma_2=s$. To make translation invariance explicit
this equation should more properly be written in terms of Nambu brackets.
However, these reduce to the above for our solution and hence we
will simplify the analysis and presentation by assuming from now on that
the equations we write apply to configurations for which $\{G_5,X^i \}=0$.

>From the generalized Nahm 
equation and the 
energy, we see that the $X^i$s can be interpreted as transverse 
coordinates to the membranes and \C{defX} represents a funnel solution
of the membranes opening into the fivebrane, while $s$ does represent the
direction along the membranes and orthogonal to the five brane. The 
classical solution \C{defX} is independent of the direction along which 
the self dual string extends in the five brane world volume. In the 
BPS limit \C{nahm}, the energy becomes
\be 
E = T_2 \int d^2 \sigma {\rm Tr} \left( 1- \frac{\lambda M_{11}^3}{8\pi} \e_{ijkl} 
\frac{d X^i}{ds} G_5 X^j X^k X^l\right),\ee    
and the energy density linearizes. The second term in this expression
is a boundary term localized on the self dual string.
In the large $N$ limit using the 
various matrix identities, this yields
\be E = N T_2 L \int ds + T_5 L \int 2\pi^2 dR R^3, \ee
where $T_5 = M_{11}^6 /(2\pi)^5$ is the five brane tension, 
and $L$ is the length of the self 
dual string. The energy has separated into two contributions--the first 
term is from the $N$ membranes while the second term is from the {\it{single}} 
five brane (both of these terms diverge because of infinite volume, but
the energy densities have the correct values). The ansatz for the membrane is 
valid at the core, but as $N \rightarrow \infty$, it matches the solution
even away from the core. The fact that the energy becomes the sum of 
the two contributions rests on the fact that the BPS condition is obeyed,
and is a consistency check for our ansatz.

>From the non--linear expression for the energy \C{energyexp} (and also 
from certain terms in the action which we shall write down later), it is 
clear that a Taylor expansion in powers of $X^i$ is possible only when
$M_{11}^6 \hat{R}^6 \ll 1$, i.e., $R \ll {\sqrt N} M_{11}^{-1}$. Thus, for 
$N \rightarrow \infty$, $R$ can be large as well. So in the large $N$ 
limit, we have a region of overlap between the five-brane and the membrane 
descriptions, which justifies matching the solutions for these theories 
as discussed below \C{sdep}.

We can simplify the expression for the energy in \C{energyexp}. We get 
that (dropping irrelevant factors)
\bea \label{energyyet} 
E = T_2 \int d^2 \sigma {\rm Tr} \bigg[ 1 + 
\left(\frac{d X^i}{ds}\right)^2
-\frac{\lambda^2}{4} 
Q^{ijk} H^{ijk} 
+\frac{\lambda^2}{16} \bigg[ \frac{d X^i}{ds} ,Q^{jkl} \bigg]
\cr \times
\bigg( \left[\frac{d X^i}{ds} ,H^{jkl} \right] 
+\left[ \frac{d X^j}{ds} 
,H^{ilk} \right] + \left[ \frac{d X^k}{ds} ,H^{ijl} \right] + \left[ 
\frac{d X^l}{ds} ,H^{ikj} \right] 
\bigg) \bigg]^{1/2}, \eea
where 
\be  Q^{ijk} = \left\{ \left[ X^i, X^j \right] ,X^k \right\},\quad 
H^{ijk} = Q^{ijk} + Q^{kij} +Q^{jki}.\ee 
Note that the projection matrix $G_5$ has dropped out of the expression.
Though it is difficult to write down the complete action for multiple
membranes, using \C{energyyet} we can write down some of the terms in 
the action. Thus we get that
\bea \label{memact} 
S = -T_2 \int d^3 \sigma {\rm Tr} \bigg[ 1 + 
\left( \p_a X^M \right)^2
-\frac{\lambda^2}{4} Q^{LMN} H^{LMN} 
+\frac{\lambda^2}{16} \left[ \p_a X^L ,Q^{MNP} \right] \cr \times
\bigg( \left[\p^a X^L ,H^{MNP} \right] 
+\left[ \p^a X^M 
,H^{LPN} \right] + \left[ \p^a X^N ,H^{LMP} \right] + \left[ 
\p^a X^P ,H^{LNM} \right] 
\bigg)  +\ldots \bigg]^{1/2}, \eea
where $X^M$ ($M=1$ to 8) are the transverse coordinates of the membranes,
$Q^{LMN} = \left\{ \left[ X^L, X^M \right] ,X^N \right\}$, 
$H^{LMN} = Q^{LMN} + Q^{NLM} +Q^{MNL}$, and $\s^a$ ($a=1$ to 3) are the 
world volume coordinates of the membranes.

\subsection{Membrane Fluctuations}

We now perform an analysis of the simplest possible fluctuations of
the membranes. This will enable us to make a consistency check that this
system indeed describes membranes ending on a five brane.
The fluctuations we consider are the overall transverse 
fluctuations of the system--these are fluctuations $\del X^m$ where $m$ 
runs over the four spatial indices transverse to both the membranes and the 
five brane. A related analysis of scattering in this system can be
found in \cite{Deger:2002zk}.

In spite of knowing only a few terms in the membrane action \C{memact}, 
we shall be able to make some statements about membrane fluctuations. 
Consistency checks with the system of D-strings ending on D3-branes
will also be helpful. For the D-brane system, suppose that we did not know 
the exact DBI action for the D-strings, but only knew some of the terms 
in the action by considering D-strings ending on a D3-brane given 
in~\cite{Constable:2001ag}. We could then try to write down a 
covariant form for these terms (as we 
have done in going from \C{energyyet} to \C{memact}), and hope to study
linearized fluctuations of the D-strings~\cite{Constable:1999ac}. Now in the 
linearized fluctuation analysis, we are essentially looking at the D-brane 
theory as dimensional reduction of ten dimensional super Yang--Mills, i.e.,
keeping only the quartic potential term alongwith the kinetic terms.    

Similarly, in our case, we need to consider a generalization of the membrane
action keeping the $O(X^6)$ potential term leading to
\be \label{memact2}
S = -T_2 \int d^3 \sigma {\rm Tr} \sqrt{ 1 + \left( \p_a X^M \right)^2
-\frac{\lambda^2}{4} Q^{LMN} H^{LMN} }. \ee
Clearly the first two terms under the square root in the action \C{memact2} 
can be obtained from the expression 
\be \label{NambuGoto}
-T_2 \int d^3 \sigma {\rm Tr} \sqrt{ -{\rm det} (P[G]_{ab})},\ee
i.e., from the determinant of the pullback of the metric to the 
worldvolume coordinates. This is because, in flat space, in static gauge,
\be \label{valP}
P[G]_{ab} = \eta_{ab} + \p_a X^M \p_b X^M ,\ee      
which gives rise to the kinetic terms in \C{memact2} when \C{valP} is
inserted in \C{NambuGoto} (this is essentially the Nambu--Goto part of the 
membrane action). So we take \C{NambuGoto} to generalize the kinetic 
terms in \C{memact2} (we shall discuss the potential term shortly). Using 
the classical solution $X^i = \hat{R} G^i$, where $\hat{R} = 1/{\sqrt s}$, we
can compute \C{NambuGoto} keeping upto terms quadratic in the fluctuations. 

We consider the fluctuation $\del X^m (t,s,\s) = f^m 
(t,s,\s) 1_{N}$, i.e., fluctuations proportional to the identity. Clearly
these are the simplest fluctuations that we can possibly analyze.
Here $\s$
is the coordinate along the self dual string. Then we get that
\be \label{nopot}
\sqrt{ -{\rm det} (P[G]_{ab})} = \sqrt{ H -H (\p_t f^m)^2 +(\p_s f^m)^2 
+H (\p_\s f^m)^2},\ee    
where 
\be H (s) = 1 +\frac{\pi N}{2 M_{11}^3 s^3}.\ee             
(This actually resembles the form of the similar terms in the action for 
the D string theory.) 

Now we consider the contribution of the potential term to the action. The 
part of it involving $Q^{ijk} H^{ijk}$ contributes to the classical 
background and can possibly change the first term in \C{nopot}, i.e., the 
coefficient of the $1/s^3$ term in $H$, but not the $H$s multiplying 
$(\p_t f^m)^2$ and $(\p_\s f^m)^2$. So this will not change the equation of 
motion for fluctuations and is irrelevant for our 
purposes.\footnote{In the D-brane 
case, there is actually no change as this occurs as an overall 
factor--presumably the same is true here, which can be checked if the 
complete membrane action is known.} Now let us turn to the contribution 
of the potential term to the fluctuations. Because $\del X^m = f^m 1_N$,
we get that (keeping only terms quadratic in the fluctuations)
\be Q^{LMN} H^{LMN} = 4 (f^m)^2 [X^i, X^j]^2 .\ee 
Thus we get that
\be \label{flucact}
S = -T_2 \int d^3 \sigma {\rm Tr} \sqrt{ H -H (\p_t f^m)^2 +(\p_s f^m)^2 
+H (\p_\s f^m)^2 -\lambda^2 (f^m)^2 [X^i, X^j]^2 }. \ee      
(The analogue of the potential term contributing to fluctuations proportional
to the identity is absent for the D string theory.) We now proceed to
evaluate $[X^i, X^j]^2 = [G^i ,G^j]^2 /s^2 = 4 (G^{ij})^2 /s^2$.

Using the definition \C{rotgen}, it is easy to show that $[(G^{ij})^2 
,G^{kl}]=0$, and using the invariance of $(G^{ij})^2$ under chirality
flip $\Gamma_5 \rightarrow -\Gamma_5$, it follows that
\be [G^i, G^j]^2 = 2 G^i G^j G^i G^j -2 N^2 (\Prp + \Prm)
\sim (\Prp + \Prm),\ee
where we have used \C{sqval}. So we have to evaluate
\be  
G^i G^j G^i G^j \Prp = \sum_{r,s,t,u=1}^n
\rho_r (\Gamma^i P_-) \rho_s (\Gamma^j P_+) \rho_t (\Gamma^i P_-) 
\rho_u (\Gamma^j P_+) \Prp. \ee  
In addition to the relations listed in \C{idmore}, we also need the 
relations
\bea 
\sum_i (\Gamma^i \otimes \Gamma^i) (P_+ \otimes P_+)_{\rm{sym}} =
\sum_i (\Gamma^i \otimes \Gamma^i) (P_- \otimes P_-)_{\rm{sym}} =0.
\eea 
Thus we obtain that
\be G^i G^j G^i G^j \Prp = (n+1)(n^2 -2n -3)\Prp, \ee
finally leading to
\be \label{quadval}
[G^i ,G^j]^2 = -\frac{(n+1)(n^3 +3n^2 +23n +21)}{2}.\ee
Inserting \C{quadval} into the action in \C{flucact} and keeping upto terms  
quadratic in the fluctuations, we obtain the equation for 
linearized fluctuations
\be \label{fluceqn}
(H \p_t^2 -\p_s^2 -H \p_\s^2 ) f^m (t,s,\s) 
+\lambda^2 \frac{(n+1)(n^3 +3n^2 +23n +21)}{2 s^2} f^m (t,s,\s) =0.\ee

Now our configuration should have two distinct interpretations in two 
distinct limits--as $s \rightarrow \infty$, it should represent 
multiple membranes while as $s \rightarrow 0$, it should represent the 
fivebrane, with the self--dual
string in its world volume. We show that this structure is consistent
with \C{fluceqn}. As $s \rightarrow \infty$, \C{fluceqn} trivially
reduces to
\be (-\p_t^2 +\p_s^2 +\p_\s^2) f^m =0, \ee    
representing free plane waves propagating in the membrane world volume 
having $SO(2,1)$ symmetry.

As in similar treatments of other brane
systems~\cite{Callan:1997kz,Constable:1999ac}, the $s \rightarrow 0$ limit
is strictly speaking outside the range of validity of our approximations, 
but nonetheless correctly matches on to the free wave equation we expect 
from the M5-brane point of view.
To see this, we note that as $s \rightarrow 0$ (i.e., 
as $R \rightarrow \infty$), the $1/s^3$ term in $H$ dominates 
and \C{fluceqn} yields
\be (-\p_t^2 + \p_\s^2) f^m + R^{-3} \frac{\p }{\p R}
\left( R^3 \frac{\p f^m}{\p R} \right) =0.\ee 
This precisely represents free plane waves propagating in the five brane 
world volume having $SO(1,1) \times SO(4)$ symmetry due to the presence of
the self--dual string.\footnote{Note that the potential term in \C{fluceqn}
drops out in both the $s \rightarrow \infty$ and the $s \rightarrow 0$ 
limits.} Thus the structure of the action, which was crucial 
in obtaining \C{fluceqn} , is consistent with the interpretation of the 
system as membranes ending on a five brane. 
 
Ideally one would like to do a general analysis of fluctuations of the 
fuzzy three sphere. However, apart from the fuzzy two sphere, it is 
difficult to do so for all other cases. For the fuzzy two sphere, the 
analysis was done by~\cite{deWit:1988ig} where generic fluctuations were 
characterized by symmetric traceless polynomials in the equivalents of the 
$G^i$s (which satisfy the Lie algebra of $SU(2)$ for the fuzzy two sphere).    
However, for all other cases, the $G^i$s by themselves do not close to 
form a Lie algebra, and one has to add other matrices transforming in 
other representations as well to make the algebra close, and generic
fluctuations involve polynomials built from all the matrices.   

\section{Discussion and Open Problems}

We have proposed an equation that describes an M2-brane ending on an M5-brane
from the M2 world-volume point of view. This generalizes Nahm's equation which
describes a D1-brane ending on a D3-brane.

In this equation the membrane degrees of freedom are represented by $N$
by $N$ matrices. On the other hand, there are various indications that $N$
membranes in M-theory have $N^{3/2}$ degrees of freedom. How can we
reconcile this with our proposal? 

The D1-brane theory is free in the ultraviolet and strongly coupled 
in the infrared.
The derivation of Nahm's equation as a BPS condition for configurations 
of the D1-brane
follows from analysis of the classical action, and hence is strictly 
speaking only valid
in the ultraviolet, that is for very short D1-branes, or equivalently 
for light monopoles
(from the D3-brane point of view). However, the non-renormalization theorems of
$N=4$ supersymmetry guarantee that many results derived from the analysis of
Nahm's equation will continue to hold even in the infrared, that is for 
heavy monopoles.

We believe that something similar may be at work in the M2-brane system. The
counting of $N^{3/2}$ degrees of 
freedom~\cite{Klebanov:1996un,Klebanov:1997kc} is a counting 
of the infrared degrees of
freedom of the theory. This same result holds for D2-branes (see e.g. the 
discussion in section 6.1 of~\cite{Aharony:1999ti}), and they 
clearly have
a UV description in terms of matrix degrees of freedom. Similarly, there may be
a description of multiple M2-branes by matrices in the UV which flows to 
a superconformal
theory in the IR with $N^{3/2}$ degrees of freedom. It is also possible 
that our
description by matrices obeying the algebra of the fuzzy
three-sphere is only some approximation to the correct description.

We have tried to reduce our system directly to the usual Nahm equation by
reducing the M2-M5 system to the D1-D3 system through compactification
and T-duality. Unfortunately we have not been able to obtain the Nahm
equation in a straightforward way. It may be that the relationship between 
the UV description of M2 and D2-branes is more subtle than a direct 
identification of the two matrix descriptions.

It would be interesting to try to use our generalization
of the Nahm equation for self--dual strings to 
study their moduli space, as has been done for monopoles. The Nahm
data also allow one to reconstruct the monopole solution on the D3-brane
world-volume. It would be interesting to see if our generalization could be
used to give some clues as to the form of the non-Abelian tensor theory
which governs M5-brane dynamics. As mentioned earlier, it would also be
interesting to explore solutions to our equation on a finite interval
which should represent finite length M2-branes suspended between M5-branes.

\section*{Acknowledgements}

We would like to thank D. Berman, D. Kutasov and S. Sethi for useful 
discussions and N. Lambert for helpful correspondence.
We would also like to thank D. Berman and N. Copland for pointing out an error 
in the earlier version of the paper. This work was  supported in part by  NSF Grant No.
PHY-0204608. 

\appendix
\section{The Fuzzy Three Sphere}
\label{fuzzy}

By analogy to the string theory 
constructions, it is natural to think that the four transverse coordinates 
to the self dual string parametrizing the five brane world volume form 
a fuzzy three sphere. The 
fuzzy three sphere has been constructed in~\cite{Guralnik:2000pb} in the 
context of solving the equations of motion for fields in the world volume 
theory of non-BPS D0 branes in a background with non-vanishing five form flux
in the IIB theory. The three sphere construction has been generalized for 
odd fuzzy spheres in~\cite{Ramgoolam:2001zx,Ramgoolam:2002wb}.      

We review and derive a few results for the fuzzy three sphere which will be 
useful later. We follow the notation 
in~\cite{Guralnik:2000pb,Ramgoolam:2001zx,Ramgoolam:2002wb}. Consider the $N 
\times N$ matrices $G^i$ ($i=1$ to 4) where $N = \frac{(n+1)(n+3)}{2}$, 
and $n$ is an odd integer. The matrices are given by
\be G^i =  \Prp \sum_{s=1}^n \rho_s (\Gamma^i P_-) \Prm + \Prm 
\sum_{s=1}^n \rho_s (\Gamma^i P_+) \Prp. \ee   
Here
\be \sum_{s=1}^n \rho_s (\Gamma^i) = (\Gamma^i \otimes \ldots \otimes 1
+\ldots +1 \otimes \ldots \otimes \Gamma^i )_{\rm{sym}}, \ee
where sym stands for the completely symmetrized $n-$fold tensor product 
representation of $spin(4)$. Also $P_{\pm} = \frac{1}{2} (1 \pm \Gamma_5)$,
and $\Prp, \Prm$ are projection operators onto the irreducible 
representations $\cRp, \cRm$ respectively of $spin(4)$. Using $spin(4)
=SU(2) \times SU(2)$, $\cRp, \cRm$ are given by the 
$(\frac{n+1}{4},\frac{n-1}{4})$ and $(\frac{n-1}{4},\frac{n+1}{4})$ 
representations respectively. Also consider the matrix $G_5$ given by
\be G_5 = \Prp - \Prm. \ee
(For $n=1$, the matrices $G^i$ and $G_5$ become $\Gamma^i$ and $\Gamma_5$ 
respectively.) So the generators of $spin(4)$ are given by
\bea \label{rotgen}
G^{ij} =  \frac{1}{2} [G^i ,G^j] = \Prp \left( \sum_{r} \rho_r 
( \Gamma^{ij} P_+) + \sum_{r \neq s} \rho_r (\Gamma^{[i} P_-) \rho_s 
(\Gamma^{j]} P_+)\right) \Prp \cr + \Prm \left( \sum_{r} 
\rho_r (\Gamma^{ij} P_-) + \sum_{r \neq s} \rho_r (\Gamma^{[i} 
P_+ ) \rho_s (\Gamma^{j]} P_- )\right) 
\Prm . \eea

In contrast to the case of even fuzzy spheres, for odd fuzzy spheres we must
deal with a reducible representation $ \cR = \cRp \otimes \cRm$. The $G^i$
are elements of $End(\cR)$. We can write $G^i = G^i_+ + G^i_-$ with
$G^i_\pm = \frac{1}{2} (1 \pm G_5) G^i$ and then $G^i_\pm$ act as homomorphisms
from $\cR_\mp$ to $\cR_\pm$.

Using the above definitions, it follows that
\be \e_{ijkl}[G_5 G^i G^j G^k G^l, G^{mn}] =0.\ee
So, $\e_{ijkl} G_5 G^i G^j G^k G^l$ is proportional to the identity 
operator in each irreducible representation, and using the symmetry under 
chirality flip ($\Gamma_5 \rightarrow -\Gamma_5$), it follows that  
$\e_{ijkl} G_5 G^i G^j G^k G^l$ is proportional to the identity operator,
i.e.
\be \e_{ijkl} G_5 G^i G^j G^k G^l \sim (\Prp + \Prm).\ee
We now proceed to calculate the proportionality constant (which is a 
function of $n$). We have that
\be \label{identity} 
\e_{ijkl} G_5 G^i G^j G^k G^l \Prp = \e_{ijkl} \sum_{r,s,t,u=1}^n
\rho_r (\Gamma^i P_-) \rho_s (\Gamma^j P_+) \rho_t (\Gamma^k P_-) 
\rho_u (\Gamma^l P_+) \Prp. \ee
The various terms in \C{identity} can be simplified using the relations 
listed below. (A similar analysis was done to evaluate $(G^i)^2$ 
in~\cite{Ramgoolam:2002wb}.) The relevant relations are given by
\bea \label{idmore} \sum_r \rho_r (P_+) \Prp = \frac{(n+1)}{2} \Prp ,
\qquad \quad \cr
\sum_{r \neq s} \rho_r (P_-) \rho_s (P_+) \Prp = \frac{(n+1)(n-1)}{4}
\Prp ,\cr
\sum_{r \neq s} \rho_r (P_+) \rho_s (P_+) \Prp = \frac{(n+1)(n-1)}{4}
\Prp ,\cr
\sum_{r \neq s \neq t} \rho_r (P_-) \rho_s (P_+) \rho_t (P_+) \Prp = \frac{(n+1)(n-1)^2}{8}
\Prp ,\cr
\sum_{r \neq s \neq t} \rho_r (P_-) \rho_s (P_-) \rho_t (P_+) \Prp = \frac{(n+1)(n-1)(n-3)}{8}
\Prp ,\cr
\sum_i (\Gamma^i \otimes \Gamma^i) (P_+ \otimes P_-)_{\rm{sym}} =
2 (P_- \otimes P_+)_{\rm{sym}},\cr
\sum_{ij} (\Gamma^{ij} \otimes \Gamma^{ji}) (P_+ \otimes P_+)_{\rm{sym}} =
4 (P_+ \otimes P_+)_{\rm{sym}},\cr
\sum_{ij} (\Gamma^i \otimes \Gamma_{ij} \otimes \Gamma^j) (P_- \otimes P_+ \otimes P_+)_{\rm{sym}} =
-2 (P_+ \otimes P_+ \otimes P_-)_{\rm{sym}},\cr
\sum_{ij} (\Gamma^i \otimes \Gamma_{ij} \otimes \Gamma^j) (P_- \otimes P_- \otimes P_+)_{\rm{sym}} =
2 (P_+ \otimes P_- \otimes P_-)_{\rm{sym}}.
\eea
Using \C{idmore}, the only relevant non-trivial relations are
\bea \label{bigrel} 
\e_{ijkl} \sum_r \rho_r (\Gamma^i \Gamma^j \Gamma^k \Gamma^l P_+) \Prp =
12 (n+1) \Prp, \cr
\e_{ijkl} \sum_{r \neq s} \rho_r (\Gamma^i \Gamma^j \Gamma^k P_-)
\rho_s (\Gamma^l P_+) \Prp = 3(n+1)(n-1) \Prp,\cr
\e_{ijkl} \sum_{r \neq s} \rho_r (\Gamma^i P_-) \rho_s (\Gamma^j \Gamma^k
\Gamma^l P_+) \Prp = 3(n+1)(n-1) \Prp,\cr
\e_{ijkl} \sum_{r \neq s} \rho_r (\Gamma^{ij} P_+) \rho_s (\Gamma^{kl} 
P_+) \Prp = 2 (n+1)(n-1) \Prp, \cr
\e_{ijkl} \sum_{r \neq s \neq t} \rho_r (\Gamma^i P_-) [\rho_s (\Gamma^j \Gamma^k P_- )
\rho_t (\Gamma^l P_+) + \rho_s (\Gamma^j P_+) \rho_t (\Gamma^k 
\Gamma^l 
P_+)]\Prp \cr
=(n-2)(n-1)(n+1) \Prp.
\eea
Plugging the expressions in \C{bigrel} into the various terms in \C{identity} 
finally gives us
\be  
\e_{ijkl} G_5 G^i G^j G^k G^l =(n+1)(n+2)(n+3) (\Prp + \Prm).\ee
Now the operators act on the space $\cR$ given by the direct sum $\cR = 
\cRp \oplus \cRm$. Let $\Pr$ be the corresponding projection operator given by
$\Pr = \Prp + \Prm$. Thus \C{idmore} leads to
\be  
\e_{ijkl} G_5 G^i G^j G^k G^l \Pr =(n+1)(n+2)(n+3) \Pr.\ee
So we write 
\be \e_{ijkl} G_5 G^i G^j G^k G^l = (n+1)(n+2)(n+3),\ee
where the right hand side means the identity operator, which we shall not be 
writing explicitly in such cases. This helps us to deduce the structure of 
the operator $\e_{ijkl} G_5 G^j G^k G^l$ which will be useful to us later. 
On general grounds, the only possible structure of this operator can be
\be \e_{ijkl} G_5 G^j G^k G^l = f(n) G^i + g(n) G_5 G^i.\ee
(Note that $g(1) =0$.) Contracting both sides with $G^i$ and using the 
relation~\cite{Ramgoolam:2001zx,Ramgoolam:2002wb}
\be \label{sqval} (G^i)^2 = \frac{(n+1)(n+3)}{2} = N, \ee
we get that
\be \left\{ f(n) + 2 (n +2) \right\} =  g(n) G_5, \ee 
from which it follows that
\be f(n) = - 2(n+2), \qquad g(n) =0,\ee
using the linear independence of $\Prp + \Prm$ and $\Prp -\Prm$. 
Thus we obtain the equation
\be \label{matrixeqn}
G^i +\frac{1}{2(n+2)} \e_{ijkl} G_5 G^j G^k G^l =0.\ee
Such a solution to the matrix equation \C{matrixeqn} has been observed
in~\cite{Sheikh-Jabbari:2004ik}. 





\providecommand{\href}[2]{#2}\begingroup\raggedright\endgroup

\end{document}